\documentclass[twocolumn,showpacs,superscriptaddress,prl,amsmath,amssymb,floatfix]{revtex4-1}
\usepackage{amsmath,amssymb}
\usepackage{bm}
\usepackage{epsfig}

\begin{document}

\title{Scaling Theory of the Mott Transition and Breakdown of the Gr\"uneisen Scaling Near a Finite-Temperature Critical End Point} 

 \author{Lorenz Bartosch}
 \affiliation{Institut f\"{u}r Theoretische Physik,
Goethe-Universit\"{a}t,
60438 Frankfurt am Main, Germany}

\author{Mariano de Souza}
\affiliation{Physikalisches Institut,
Goethe-Universit\"{a}t,
60438 Frankfurt am Main, Germany}

\author{Michael Lang}
\affiliation{Physikalisches Institut,
Goethe-Universit\"{a}t,
60438 Frankfurt am Main, Germany}

\date{April 27, 2010}

 \begin{abstract}
{
We discuss a scaling theory of the lattice response in the vicinity of a finite-temperature critical end point. The thermal expansivity is shown to be more singular than the specific heat such that the Gr\"uneisen ratio diverges as the critical point is approached, except for its immediate vicinity. More generally, we express the thermal expansivity in terms of a scaling function which we explicitly evaluate for the two-dimensional Ising universality class. Recent thermal expansivity measurements on the layered organic conductor $\kappa$-$(\text{BEDT-TTF})_2 X$ close to the Mott transition are well described by our theory.
}
\end{abstract}


\maketitle

Scale invariance and universality
are among the most striking concepts in physics \cite{Fisher83, Goldenfeld92, Cardy96, Kardar07}. 
A prominent example is the critical end point of the liquid--gas transition of water at which the correlation length diverges and fluctuations occur on all length scales.
Even though quantum mechanics can be important for understanding the nature of the different phases present, in the vicinity of a finite-temperature critical end point, a purely classical theory suffices to describe the transitions between these phases.
As is well-known from the scaling theory of critical phenomena, every critical end point is characterized by a universal scaling function and a set of critical exponents, known as the scaling exponents \cite{Fisher83, Goldenfeld92, Cardy96, Kardar07}. 
The scaling function as well as the scaling exponents do not depend on microscopic details of the underlying system. Instead, they are entirely determined by a few general properties like the dimensionality of the system, the symmetry properties of the order parameter, or the presence or absence of long-ranged interactions.

A very interesting class of materials consists of the so-called Mott insulators which have an odd number of conduction electrons per unit cell and, according to band theory, should be metals. However, if the ratio of the on-site correlation energy $U$ to the kinetic energy $W$ becomes larger than a critical value, all electrons localize. Prominent examples for the Mott transition are the 
Cr-doped 
$\text{V}_2 \text{O}_3$ 
\cite{Limelette03} and the layered organic charge-transfer salts $\kappa$-$(\text{BEDT-TTF})_2 X$ \cite{Lefebvre00,Limelette03b,Fournier03,Kagawa05,deSouza07,Toyota07,Kagawa09}. Here, BEDT-TTF is bis(ethylenedithio)tetrathiafulvalene, hereafter abbreviated ET, and $X$ stands for various kinds of monovalent anions.
\begin{figure}[tb]
  \includegraphics*[width=0.8\columnwidth]{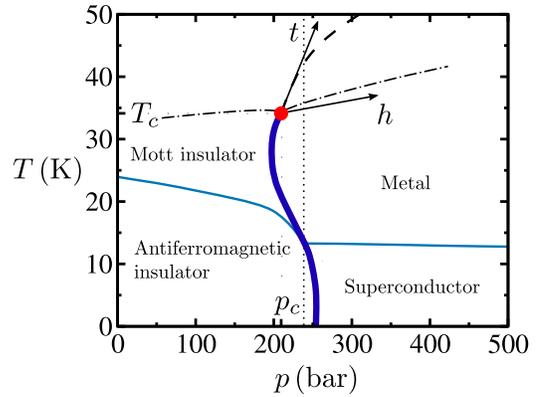}
  \caption{%
(Color online) Simplified phase diagram of the $\kappa$-$(\text{ET})_2 X$ family as mapped out in Refs.~\cite{Lefebvre00,Limelette03b,Fournier03}. While the salt with $X=\text{Cu}[\text{N} (\text{CN})_2]\text{Cl}$ (located at $p=0$) is a Mott insulator,
the hydrogenated variant of the $X=\text{Cu}[\text{N} (\text{CN})_2]\text{Br}$ salt  (\emph{h}8-Br) is a superconductor. 
Its deuterated variant (\emph{d}8-Br), investigated in Ref.~\cite{deSouza07}, is located at ambient pressure (dotted line) very close to the critical region.
For simplicity, regions of inhomogeneous phase coexistence 
are left out.
The thick, solid line corresponds to a first-order transition. It terminates in a critical end point (red dot), above which there is only a crossover between the Mott insulator and the metal (dashed line).
The crossover line differs from the dashed-dotted line $t\approx 1.74 |h|^{8/15}$ where signatures in the lattice response are expected to be the strongest.
}
  \label{fig:phasediagram}
\end{figure}
As 
shown in Fig.~\ref{fig:phasediagram}, their ground states can be tuned by applying either chemical or external pressure.
While simple arguments based on symmetry \cite{Castellani79} suggest that the Mott transition lies in the Ising universality class \cite{Kotliar00,Onoda03,Georges04,Papanikolaou08}, this scenario is by no means obvious and is still under debate \cite{ 
Imada05,Imada10,Dzero10}.
Using the pressure-sweep technique to control the ratio $U/W$, conductivity measurements by Limelette {\em et al.}\ on 
$(\text{V}_{1-x} \text{Cr}_x)_2 \text{O}_3$ 
near the critical end point
were reported to reveal mean-field behavior, with a crossover to
scaling properties with $3D$ Ising critical exponents \cite{Limelette03}.
However, \emph{unconventional critical behavior} was claimed by Kagawa {\em et al.}\ for pressurized quasi-$2D$ $\kappa$-$(\text{ET})_2 \text{Cu}[\text{N}(\text{CN})_2]\text{Cl}$ (abbreviated $\kappa$-Cl) \cite{Kagawa05}.
The critical exponents obtained from the conductivity measurements obeyed scaling relations but did not match any established universality class.
In particular, Ising criticality was ruled out by the authors.

A different explanation of the experiments by Kagawa {\em et al.}\ was given by Papanikolaou {\em et al.}~\cite{Papanikolaou08}, who pointed out that the conductivity can have a different scaling behavior than the order parameter $m$ itself. In particular, for the $\kappa$-$(\text{ET})_2 X$ family, it was argued that 
the conductivity $\Sigma$ should scale like the energy density, implying $\delta \Sigma \propto |m|^\theta$, where $\theta = (1-\alpha)/\beta$ is a combination of the critical exponents $\alpha$ and $\beta$ related to the singular behavior of the specific heat and the order parameter, respectively. With $\theta = 8$ for the $2D$ Ising universality class, it was shown that the conductivity measurements in Ref.~\cite{Kagawa05} are actually consistent with the assumption of Ising criticality.
A complementary experiment, which concerns the anomalous response of the lattice in the vicinity of the Mott transition of deuterated $\kappa$-$(\text{ET})_2 \text{Cu}[\text{N}(\text{CN})_2]\text{Br}$, was reported in Ref.~\cite{deSouza07}.
The data, which were analyzed by assuming Gr\"uneisen scaling
between the thermal expansivity $\alpha_p(T)$ and the specific heat $c_p(T)$ 
[i.e., $\alpha_p(T) = \Gamma \, c_p(T)$], were found to be at odds with the experiment by Kagawa {\em et al.} \cite{Kagawa05}.
As the expansivity measurements were performed very close to but not exactly 
at the critical point observed in Ref.~\cite{Kagawa05}, a definition of a critical exponent is problematic.
In this Letter, 
we demonstrate the breakdown of Gr\"uneisen scaling 
in the vicinity of a finite-temperature critical end point.
By elaborating on a scaling theory of the expansivity 
we reanalyze the expansivity data  of $\kappa$-$(\text{ET})_2 X$ and find consistency with $2D$ Ising criticality.

By identifying the pressure $p$ and the temperature $T$ as independent parameters to fine-tune the phase transition, we can already infer that the change of volume $\Delta V$ at the Mott transition is proportional to the order parameter $m$. This is a simple consequence of the fact that the volume $V$ and pressure $p$ are conjugate variables, and it implies that the change in volume across the Mott transition 
should scale as
$\Delta V \propto (T_c -T)^\beta$, where $\beta=1/8$ for the $2D$ Ising universality class.
Apart from the different value of the critical exponent $\beta$, this is quite analogous to the familiar liquid--gas transition.

To set up a complete scaling theory, let us consider the singular part of the Gibbs free energy,
$f(t,h)$, which,
for the $2D$ Ising universality class, can be written in the scaling form (see e.g., Ref.~\cite{Fonseca03})
\begin{equation}
  \label{eq:fIsing}
  f(t,h) = \frac{t^2}{8\pi} \ln t^2 + |h|^{d/y_h} \Phi \left( t/|h|^{y_t/y_h} \right) .
\end{equation}
Here, $t=(T-T_c- \zeta (p-p_c))/T_0$ and $h=(p-p_c- \lambda (T-T_c))/p_0$ are 
temperature-like and pressure-like scaling variables which also contain small linear mixing terms. 
Apart from the critical temperature $T_c$ and pressure $p_c$, the scaling variables depend on the four nonuniversal constants $\lambda$, $\zeta$, $T_0$, and $p_0$.
The finite values of $\lambda$ and $\zeta$ account for the tilted scaling axes shown in Fig.~\ref{fig:phasediagram}. For example, $\lambda = dp/dT$ can be identified as the value of the finite slope at the end of the first-order transition line.
Let us for now assume that in suitable units the parameters $\lambda$ and $\zeta$ are sufficiently small such that it is reasonable to use $\partial t / \partial p \approx 0$ and $\partial h /\partial T \approx 0$. 
As further explained below, this reasoning can be justified except for the nearest vicinity of the critical point.

\begin{figure}[b]
\centering
  {\includegraphics*[width=0.8\columnwidth]{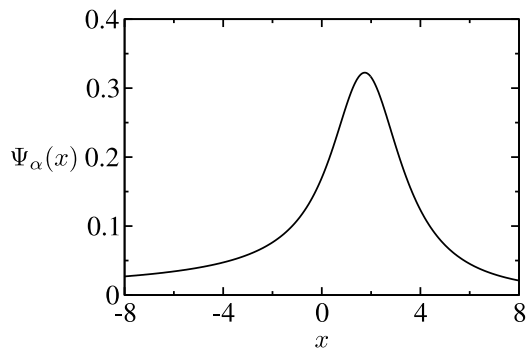} \quad}
  \caption{%
Plot of the scaling function $\Psi_\alpha(x)$ for the $2D$ Ising universality class. 
}
  \label{fig:scalingfunction}
\end{figure}

\begin{figure}[tb]
  \includegraphics*[width=0.8\columnwidth]{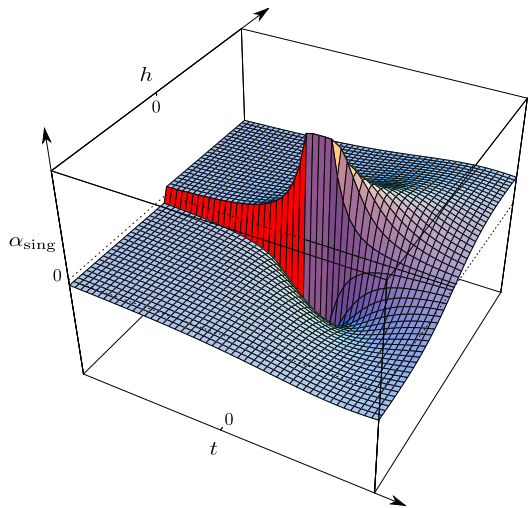}
  \caption{%
(Color online) Plot of the thermal expansivity $\alpha_{\text{sing}}$ as a function of $t$ and $h$ which diverges at the critical end point $t = h = 0$.
The line of first-order phase transitions along the negative $t$-axis [where the volume of the system changes according to $\Delta V \propto (-t)^{1/8}$] is clearly accompanied by a branch cut in the thermal expansivity.
Note that due to the scale invariance near the critical point this plot looks the same on all length scales, i.e. rescaling $t \to b^{y_t} t$ and $h \to b^{y_h} h$ with arbitrary $b$ we obtain exactly the same graph.}
  \label{fig:3dplot}
\end{figure}

While the renormalization group eigenvalues $y_t = 1$ and $y_h = 15/8$ for the $2D$ Ising universality class are known from Onsager's famous solution, the scaling function $\Phi(x)$ is not known analytically and has to be evaluated numerically. This, however, can be done with very high precision \cite{Fonseca03, Mangazeev09}.
All thermodynamic critical exponents can be derived by taking derivatives of Eq.~(\ref{eq:fIsing}) with respect to $t$ or $h$. For example, if $V$, $S$, and $G$ denote the volume, the entropy, and the Gibbs free energy of the system, respectively, the specific heat is given by
\begin{equation}
  c_p  = \frac{T}{V} \left(\frac{\partial S}{\partial T}\right)_{p}
  = - \frac{T}{V} \left(\frac{\partial^2 G}{\partial T^2}\right)_{p} .
  \label{eq:specificheat}
\end{equation}
For $\lambda = h=0$ this reduces to the well-known result for the singular part of the specific heat,
\begin{equation}
c_{\text{sing}} \propto 
- \left(\frac{\partial^2 f}{\partial t^2}
\right)_{h=0}
\propto - \ln | t | ,
\end{equation}
which, due to the absence of a power of $t$, implies a critical exponent $\alpha = 0$.
Of central importance here is the singular part of the thermal expansivity $\alpha_p = 
V^{-1} \left({\partial V}/{\partial T}\right)_p =
  - V^{-1} \left({\partial S}/{\partial p}\right)_T$,
\begin{equation}
  \label{eq:expansivity}
  \alpha_{\text{sing}} \propto \frac{\partial^2 f}{\partial h\, \partial t} ,
\end{equation}
which for $h \to \pm 0$ and $t < 0$ scales as (see also Ref.~\cite{Papanikolaou08})
\begin{equation}
  \label{eq:expansivitypc}
   \alpha_{\text{sing}} \Big|_{h= \pm 0} \propto \text{sgn} (h)\, (-t)^{-1+\beta} . 
\end{equation}
The generalized Gr\"uneisen ratio $\Gamma_p(T)$ is defined as 
\begin{equation}
  \label{eq:Grueneisen}
  \Gamma_p (T)= \frac{\alpha_{p}(T)}{c_p(T)} = - \frac{1}{T}
  \frac{\left( \partial S / \partial p \right)_T}{\left( \partial S / \partial T \right)_p} .
\end{equation}
In a different situation, where there is a line of second-order transitions (but no critical end point), the pressure $p$ enters only via the critical temperature $T_c(p)$, such that taking derivatives with respect to $p$ is essentially the same as taking derivatives with respect to $T$. As a consequence, the Gr\"uneisen ratio stays constant even if $c_p(T)$ and $\alpha_p(T)$ diverge.
However, this is quite different at a critical end point where $t$ and $h$ are independent parameters.
While it is, by now, well established that the Gr\"uneisen ratio can diverge at a quantum critical point \cite{Zhu03}, a strong increase in $\Gamma_p(T)$ is also found for a critical end point at finite temperature, where for $t<0$ (and $\lambda = 0$)
\begin{equation}
  \label{eq:Grueneisencriticalendpoint}
  \Gamma_{\text{sing}} = \left. \frac{\alpha_{\text{sing}}}{c_{\text{sing}}} \right|_{h=\pm 0} \propto \text{sgn}\,(h) \, (-t)^{-1+\alpha+\beta} .
\end{equation}
It should also be noted that for $h=0$ and $t>0$ the singular part of the Gr\"uneisen ratio vanishes. 
A more complete picture of the vicinity of the critical point can be derived from the scaling function.
To this end, we combine Eqs.~(\ref{eq:fIsing}) and (\ref{eq:expansivity}) to obtain the scaling form of the expansivity,
\begin{equation}
  \label{eq:expansivityoftandh}
  \alpha_{\text{sing}}(t,h) \propto \text{sgn} (h)\, |h|^{-1+(d-y_t)/y_h} \Psi_\alpha(t/|h|^{y_t/y_h}) .
\end{equation}
Here, $d$ is the dimension of the system.  The scaling function $\Psi_\alpha(x)$ 
is given by
\begin{equation}
  \label{eq:Psi}
  \Psi_\alpha(x) = \frac{d-y_t}{y_h} \Phi'(x) -\frac{y_t}{y_h} x \Phi''(x).
\end{equation}
A plot of 
$\Psi_\alpha(x)$ for the $2D$ Ising universality class is shown in Fig.~\ref{fig:scalingfunction}
\footnote{We have calculated the scaling function $\Phi(x)$ and its derivatives numerically using the methods outlined in Section~8 of Ref.~\cite{Fonseca03}. It should be noted that to obtain accurate data for $1 \lesssim |x| \lesssim 3$ it is not sufficient to use the series expansions of the scaling function $\Phi(x)$ discussed in Refs.~\cite{Fonseca03, Mangazeev09}
}.
$\Psi_\alpha(x)$ assumes its maximum not at $x=0$, but at the finite value $x \approx 1.74$, such that for fixed $h$ the strongest lattice response is expected for $t \approx 1.74 |h|^{8/15}$, 
which is indicated by the dashed-dotted line in Fig.~\ref{fig:phasediagram}.
Using the asymptotic behavior of $\Psi_\alpha(x)$ for large values of $|x|$,
\begin{equation}
  \label{eq:Psiasymptotic}
  \Psi_\alpha(x) \propto \left\{
    \begin{array}{ll}
      (-x)^{-7/8} , & x \to -\infty , \\
      x^{-15/8} , & x \to \infty ,
    \end{array}
\right.
\end{equation}
we recover Eq.~(\ref{eq:expansivitypc}) and $\alpha_{\text{sing}} = 0$ for $h=0$
and $t>0$, 
as expected.
More generally, the singular part of the expansivity $\alpha_{\text{sing}}$ can be calculated for arbitrary values of $t$ and $h$, 
see Fig.~\ref{fig:3dplot}.
The strong lattice response at $t \approx 1.74 |h|^{8/15}$ and the sign change of the expansivity across the $h=0$ line can clearly be seen.

\begin{figure}[tb]
  \includegraphics*[width=0.8\columnwidth]{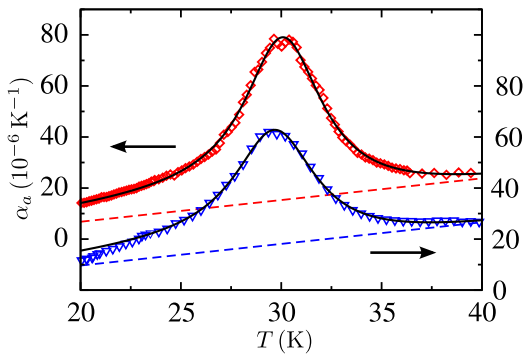}
  \caption{%
(Color online) Fit to the expansivity data for $d$8-Br crystal \#~1 (diamonds $\Diamond$) and \#~3 (triangles $\nabla$) from Ref.~\cite{deSouza07} by Eq.~(\ref{eq:fitformula}).
The fits turn out to be only weakly sensitive to the parameter $\lambda$. A least-squares fit gives, for the fitting parameters of crystal \#~1, $T_c + \zeta(p-p_c) = 27.5\,$K, $(p-p_c)/\lambda = 26.7\,$K, $A/\lambda^{7/15} = 874 \cdot 10^{-6}\,$K${}^{-8/15}$, $B / \lambda^{8/15} =3.88\,$K${}^{-7/15}$, $C = 13.2\cdot 10^{-6}\,$K${}^{-1}$, and $D = 0.85\cdot 10^{-6}\,$K${}^{-2}$.
The straight dashed lines denote the linear background contribution.
%
%
%
%
%
}
  \label{fig:fit}
\end{figure}

Having established a general scaling theory for the singular part of the thermal expansivity, we now analyze the thermal expansivity measurements of Ref.~\cite{deSouza07}.
For simplicity, we assume that the nonsingular background contribution to the thermal expansivity can be approximated by a linear term, such that for the $2D$ Ising universality class we have
\begin{align}
  \label{eq:fitformula}
  \alpha (T,p) & =  A\, \text{sgn} \left[ p-p_c  -\lambda (T-T_c) \right] \nonumber \\
& \ {} \times
\left[ p-p_c  -\lambda (T-T_c) \right]^{-7/15} \nonumber \\
& \ {} \times
  \Psi_\alpha \left(\frac{B [T-T_c - \zeta(p-p_c)]}{\left[ p-p_c - \lambda (T-T_c) \right]^{8/15}} \right)\nonumber \\
 & \ {} + C + D \, [T-T_c - \zeta(p-p_c)] .
\end{align}
For other universality classes, the exponents $7/15$ and $8/15$ should be replaced by $(1-\beta)/(\beta+\gamma)$ and $1/(\beta + \gamma)$, respectively, where $\gamma$ is the critical exponent associated with the compressibility $\kappa$.
A fit to the data published in Ref.~\cite{deSouza07} by Eq.~(\ref{eq:fitformula})
is shown in Fig.~\ref{fig:fit}.
The data are nicely described by our theory which demonstrates that the expansivity measurements of Ref.~\cite{deSouza07} are in fact consistent with the assumption that the Mott transition in the $\kappa$-$(\text{ET})_2 X$ family is in the $2D$ Ising universality class.

Finally, let us explore the linear mixing terms 
in some more detail. For finite $\lambda$, the specific heat $c_p$ and the thermal expansivity $\alpha_p$ contain terms proportional to $(\lambda T_0/p_0)^2 \kappa$ and $(\lambda T_0/p_0) \kappa$, respectively, with $\kappa$ being more singular than $c_{\text{sing}}$ and $\alpha_{\text{sing}}$ computed above.
Consequently, even if  $(\lambda T_0/p_0)$ is small, there is a region where this term becomes important and the Gr\"uneisen ratio saturates. 
A simple analysis shows that the crossover scales for the $2D$ Ising universality class are $|T-T_c| \approx \lambda^{8/7}/B^{15/7}$ and $| p -p_c | \approx (\lambda/B)^{15/7}$.
However, since for the experiment in Ref.~\cite{deSouza07} $p-p_c \approx 50\,$bar $\gg (\lambda/B)^{15/7} \approx 0.1\,$bar, 
having neglected the term proportional to $\lambda \kappa$ in our analysis above is clearly justified.
It has also been argued that a finite
shear modulus in solids can lead to a suppression of critical
fluctuations near the critical point if the order parameter is
proportional to $\Delta V$ (see Ref.\ \cite{Dzero10} and references therein).

In summary, we have developed a scaling theory for describing the
singular part of the thermodynamic expansivity in the vicinity of a
finite-temperature critical end point. As a salient result of our
study, we find for $h= \pm 0$ and $t<0$ a universally diverging Gr\"{u}neisen ratio $\Gamma_{\text{sing}} =
\alpha_{\text{sing}}/c_{\text{sing}} \propto \text{sgn}(h) (-t)^{-1+\alpha+\beta}$, except for
the immediate vicinity $|T - T_c| \lesssim (\lambda
/B^{\beta+\gamma})^{1/(\beta+\gamma -1)}$, where  $\Gamma_p(T)$ is
expected to saturate. For the Mott transition in the vicinity of the critical end point,
volume changes are a natural consequence of thermodynamics and are similar
to the liquid--gas transition, implying $\Delta V \propto (T_c - T)^\beta$.
By explicitly evaluating the scaling function for the $2D$ Ising
universality class, an excellent description of thermal expansion
data for the layered organic conductor $\kappa$-(BEDT-TTF)$_{2}X$
has been achieved, consistent with the conjecture made by the authors of Ref.~\cite{Papanikolaou08}
and their analysis of the conductivity measurements in Ref.~\cite{Kagawa05}. We
stress that the approach presented here does not rely on any assumption regarding
the system's underlying microscopic interactions, including a
possible involvement of lattice degrees of freedom \cite{deSouza07} or system
anisotropies. Our theory also predicts large lattice effects
at $t \sim |h|^{8/15}$, reaching out to both the low- and
high-pressure sides of the critical end point. 
The observations of a large anomaly in the thermal expansion of \emph{h}8-Br \cite{Lang03}, as 
compared to a tiny feature in the specific heat \cite{Nakazawa96},
and a pronounced dip in the elastic constant of pressurized
$\kappa$-Cl on both sides of the critical end point \cite{Fournier03} are
consistent with the latter prediction. In fact, by extending our
scaling theory to the singular part of the compressibility,
$\kappa_{\text{sing}} \propto \partial^{2}f/\partial h^{2}$, we find that the dip in the sound
velocity observed in Ref.~\cite{Fournier03} should follow a line close to the
dashed-dotted line in Fig.~\ref{fig:phasediagram}. Note that these lines should not be
confused with the crossover line obtained from the conductivity measurements
of Ref.~\cite{Kagawa05}. 
Finally, let us emphasize that due to its universal character, 
our scaling theory should apply to a
broad range of systems, including e.g.~metamagnetic materials, for
which an anomalously large Gr\"{u}neisen ratio and a sign change of
the expansivity have recently been reported \cite{Weickert10}.

\acknowledgments
We would like to thank C.~Gros, P.~Kopietz,  J.~M\"uller, and J.~Schmalian for valuable discussions.
This work was supported by the DFG via 
SFB/TRR 49.

\end{document}